\documentclass[11pt]{article}

\usepackage{amsfonts,amsmath,amssymb,graphics}

\newcommand{\mycaption}[1]{\caption{ \small #1 } }

\newcommand{\ex}{\mathbf{E}}

\newcommand{\rhat}{\widehat{r}}
\newcommand{\shat}{\widehat{s}}
\newcommand{\sbar}{\overline{s}}
\newcommand{\recov}{\Re}
\newcommand{\Phat}{\widehat{P}}

\newcommand{\notthis}[1]{}

\newcommand{\inv}{^{-1}}

\newcommand{\half}{\frac{1}{2}}

\begin{document}

\title{\bf The credit spread curve\\ \Large I: Fundamental concepts, fitting, par-adjusted spread, and expected return}
\author{Richard J. Martin\footnote{Dept.\ of Mathematics, Imperial College London, South Kensington, London SW1 2AZ, UK. Email: {\tt richard.martin1@imperial.ac.uk}}}
\maketitle

\begin{abstract}

The notion of a credit spread curve is fundamental in fixed income investing, but in practice it is not `given' and needs to be constructed from bond prices either for a particular issuer, or for a sector rating-by-rating. Rather than attempting to fit spreads---and as we discuss here, the Z-spread is unsuitable---we fit parametrised survival curves.
By deriving a valuation formula for a risky bond, we explain and avoid the problem that bonds with a high dollar price trade at a higher yield or spread than those with low dollar price (at the same maturity point), even though they do not necessarily offer better value. In fact, a concise treatment of this effect is elusive, and much of the academic literature on risky bond pricing, including a well-known paper by Duffie and Singleton (1997), is fundamentally incorrect.

We then proceed to show how to calculate carry, rolldown and relative value for bonds/CDS. Also, once curve construction has been programmed and automated we can run it historically and assess the way a curve has moved over time. This provides the necessary grounding for econometric and arbitrage-free models of curve dynamics, which will be pursued in later work, as well as assessing how the perceived relative value of a particular instrument varies over time. 

This version discusses CDS curve stripping in more depth, provides an extension to the previous forward hazard rate model, extends the discussion on EM bonds, 
and has new sections on accreting bonds and on bond forwards.

\end{abstract}


\section*{Introduction}

Credit investors need to answer a variety of questions about the bond markets in which they operate, such as:

\begin{itemize}

\item[A1]
How has a particular subset of the universe, e.g.\ BBB miners in the 5--10y maturity bucket, traded over the last few years? Where is it now relative to history?

\item[A2]
Is the BBB mining curve flat or steep today by comparison with history?

\item[A3]
How have A/BBB vs BB/B miners traded in the last few years? The effect of investment grade (IG) and high yield (HY) moving oppositely is often called compression/decompression.

\item[A4]
What is the influence of country spread on corporate spread, sector by sector (cf.\ \cite{Martin18c})?

\item[B1]
How has the credit spread of a particular issuer varied over an extended period of time (several years)?

\item[B2]
Has the curve of a particular issuer flattened or steepened of late?

\item[C]
How do we determine carry, rolldown and relative value of a bond? These are essential ingredients in understanding how to evaluate expected return.

\end{itemize}

Questions marked `A' are at broad sector-level and are essentially matters of data aggregation, fitting or parametrisation, taking into account rating or maturity. They might therefore seem rather trivial: why cannot we simply take a set of bonds with the desired characteristics---rating, maturity range, etc.---and then compute the average spread (weighting by issue size and duration would be common practice)? The problem with this is that over time bonds enter and exit the bucket, for various reasons: new bonds being issued; existing ones being redeemed or called; ratings changing; maturity steadily declining so that the bond moves into or out of the desired maturity range. This will create jumps in the average spread. Also if the desired bucket is too narrowly defined, e.g.\ 5--7y miners rated BB$-$, we might find no bonds at all. Finally, there is no law that states that bonds should trade monotonically with credit rating, and in practice we can easily find examples where a BBB name trades tighter than a BBB+ one (at the same maturity point), as the market spread is in a sense current, and the rating may be out of date, or perceived as such. All this indicates that we need to produce a parametrically-defined grid of curves that are of a sensible shape and vary monotonically with rating (i.e.\ do not cross). Then the estimated spread for a particular rating and tenor is a sort of weighted average of bonds of nearby rating and tenor.

By contrast `B' refers to specific issuers. In CDS markets this question seems trivial because a CDS curve is fundamental to the asset class; but in practice liquidity is concentrated at the 5y point, so that at other maturities we only have indication levels used for marking traders' books, and is nonexistent beyond 10y. In fact, most issuers do not trade in CDS, and then one must refer to a particular bond: then, the difficulty is that the bond ages over time and so part of its spread variation results from reduction in maturity. In fact this calls for the same solution as above, but with the simplification that we only want one curve, not one for each rating. While some issuers come to market frequently, giving an almost continuous picture of the term structure, others may only have one or two bonds outstanding.

This leads neatly to `C', which refers to the analysis of a particular bond. It may be helpful to define our terms:
\begin{itemize}
\item
Carry is the profit and loss (PL) contingent on the yield of the bond remaining fixed. For a par bond this is synonymous with the coupon, but for a bond trading $>100$ the bond price will decrease and for one trading $<100$ it will increase as a result of the pull-to-par-effect. The carry arises from two sources: a pure interest-rate component and a credit spread component.

\item
Rolldown is the PL contingent on yield change that arises a result of maturity reduction, with the curve assumed to remain fixed. Usually curves are upward-sloping and so the yield reduces, and the rolldown is positive, but when the curve is inverted the effect will be negative.
Carry and rolldown are often taken together, and represent the total PL arising from ageing of the bond while the curve remains fixed.

\item 
Credit relative value (RV) expresses, in spread or price terms, the degree to which a bond offers good value relative to its peers. This is often referred to as `rich/cheap to the curve'---necessitating, of course, a curve.

\end{itemize}
At the risk of repetition: rolldown and RV require the construction of an issuer's curve. (For carry it is less obvious, but there is a subtlety that we will come to presently.) But this is in practice a mythical beast, and needs to be built from available bond data.

Despite the fundamental nature of this subject, the literature, which stretches back some thirty years, is disappointing. At one end of the spectrum are highly academic treatises such as \cite{Duffie97} which develop the subject from the perspective of complete markets and martingale pricing. Despite its impression of intellectual depth, this paper is faulty in several respects. One objection is that credit markets lack the necessary completeness for the mathematical grounding to be valid, but worse than that is that the entire development follows from their incorrect eq.(1), pricing a contingent claim in terms of a `default-adjusted discount rate' $R=r+hL$, with $r$ the riskfree short rate, $h$ the hazard rate and $L$ the LGD\footnote{Loss given default, here as expressed as one minus recovery, $1-\recov$.}. The incorrectness of this stems from the fact that the loss mechanism is wrong: the discount factor is the exponential-integral of $-R_t$, which depends nonlinearly on $L$. The financial interpretation of terms in $L^2$ and higher (in the Taylor expansion around $L=0$) is that default may occur more than once in the given time interval, and that losses are in proportion to the market value of the asset just before default. But this bears no resemblance to reality: a bond can only default once, and the loss is a proportion of the par value, as that is the bankruptcy claim. This error is fundamental because the coupon stream and principal payment are different claims and require different discounting, and as we show later there is a tendency for premium bonds to trade at a higher yield than discount bonds\footnote{A bond is said to be discount if its dollar price is below par, premium if above.} of the same maturity---we call this the par/non-par problem from now on. The authors clearly do not recognise this, as they state at the beginning of their \S4, ``The inability to separately identify [the hazard rate and LGD] using defaultable bond yields \ldots'' which as we have said is at variance with market pricing. Indeed, typical of this kind of discussion (and there are plenty of other examples from academic institutions) is its lack of connection to real-world examples from the debt markets. Lando \cite{Lando98} makes more progress, to the extent of valuing, at least in theory, the coupon stream and the recoverable portion of the principal when default can occur at an unknown time, and goes on to discuss the simultaneous fitting of multiple rating curves, and also rating dynamics via a transition matrix. The par/non-par problem is not addressed, though, and there are theoretical and practical difficulties in fitting a transition matrix to bond spreads. Many different matrices can give almost identical term structure \cite{Martin21a}, and resolution of this ambiguity requires spread volatility information; and the process is quite computationally intensive. We think a simpler and quicker idea is to write down directly a parametric form for the survival curve in each rating state: that way, it is clear how the parameters directly relate to the term structure and hence to the calibration data---which is not at all the case with rating transition models.

One of the more ridiculous discussions, fortunately confined to the academic literature, is the nature of the recovery assumption. A discussion in \cite[4.2]{Berd11} points to three possibilities: (i) fractional recovery of Treasury (FRT); (ii) fractional recovery of market value just before default (FRMV); (iii) fractional recovery of par (FRP).
The only sensible course of action is (iii), because that is what the bankruptcy claim is---though the reader should pay attention to our later discussion on accreting bonds, where FRP is wrong. The only justification for (i),(ii) seems to be that they give rise to closed-form solutions in some types of model: to us, that is an entirely specious argument. That (ii) is a silly idea can be seen from the fact that at default the bond price may not even move (in fact it might increase) if erosion of value had occurred a long while beforehand and the market knew that default was imminent\footnote{Examples are OI Telecom and Samarco Minera\c{c}\~{a}o, both in 2016.}.

At the other end of the spectrum lies the `applied' literature, some of which is little more than a plug for commercial software implementations, e.g.~\cite{Deventer12} which does little to address the main issues. The `middle ground' should be occupied by articles authored by quants with a solid technical grounding and good experience of working in the markets \cite{Berd04,Berd11}; these are probably the best sources for the working quantitative analyst, but we regard the development of this paper as covering more ground more rapidly.

Here in basic terms is what we regard as the correct approach:
\begin{itemize}
\item
We should discount cash-flows using riskfree discount factors, survival probabilities, and recoveries. Methods such as Z-spread are unsuited to bonds trading away from par, but in practice the effect of recovery---which goes a long way to explain why high dollar price bonds trade at higher spread---has not been carefully dealt with. Also, no bond spread definition is compatible with CDS spread. Berd et al.\ \cite{Berd04} come to the same conclusion in an imprecise and roundabout way. Van~Deventer \cite{Deventer12} uses a maturity-smoothed version of the Z-spread, and declares that it has been successfully used since 1993: presumably he considers this to be some sort of recommendation, but instead it indicates a basic lack of understanding. In fact, Figures~\ref{fig:colom},\ref{fig:colom2} illustrate why such an approach is poor: the term structure would end up with a large kink at the 17y point, without economic justification as it is just a par/non-par effect.
\item
We do not need hazard rate models, and the existence of the survival function $Q(T)$ does not need pseudo-academic grounding in arbitrage pricing theory (which is irrelevant to the problems at hand). Instead we simply declare that the PV of a risky and unrecoverable dollar occurring time $T$ from now is the product of a riskfree discount factor $B(T)$ and the survival function, hence  $B(T)Q(T)$. The formulation for $Q$ in terms of the hazard rate $h$, i.e.\ $Q(T)=\ex_0[e^{-\int_0^T h_s \, ds}]$, is unnecessary, for it raises questions about the nature of the process $\{h_t\}$, and need not be introduced in the first place. 
\item
It is necessary to parametrise $Q(T)$ using an appropriate monotone-decreasing function and while some kind of spline is possible \cite[\S5]{Berd11} we use eq.(\ref{eq:Q1}). This is done when we have one issuer and many bonds/CDS. It is the first step in the so-called HJM approach \cite[App.C]{DuffieSingleton03}; the second is to write down the risk-neutral dynamics, which we consider in forthcoming work.
\item
When there are too few bonds to fit a curve for an issuer, or where we want to think about relative value, we index credit quality using ratings. We can then think about how a name trades relative to its rating curve, and can use either the public rating or our opinion of it, or conversely find its market-implied rating by seeing which curve prices it most closely. 
When fitting multiple rating curves we need to do them all at once, which necessitates a more general functional form for $Q(T)$, eq.(\ref{eq:Q2}) et seq., and we must be prepared for data that are very `noisy'.
\item
In principle one can PV a bond/CDS without any computational short-cuts by discounting the exact cash-flows using their exact dates. Our methods can be used this way. In practice, though, one loses little and gains a much simpler implementation by assuming the coupons to be continuously-paid\footnote{van~Deventer \cite{Deventer12} makes a fuss about using exact rather than scheduled payment dates, but a moment's thought shows that a day or two's discounting gives rise to a change in value that is negligible compared with the bid-offer of the bond.}. For CDS this is not a problem because the coupon is always 1\% or 5\% and paid quarterly. In bond markets one does occasionally see some hoary specimens with coupons $>10\%$, and when the issuer is distressed this assumption starts to be questionable. However, in such cases the survival curve is likely to be highly idiosyncratic and determined by the precise timings of the firm's cashflows, so a generic model is of limited value.
\item
As intimated above we fit to prices (or CDS PVs). However, we want to plot the spread vs maturity and this necessitates what we are calling the par-adjusted spread, which removes the par/non-par effect.
\end{itemize}

\section{Methods}
\label{sec:quant}

The credit-riskiness of a bond is, for any performing credit, best encapsulated by a quantity known as the \emph{spread} which, loosely, indicates how much yield it has by comparison with a riskfree bond of the same maturity. So this is what we are trying to model.
(When a credit is distressed and unlikely to make its payments, its yield becomes so high as to be meaningless and then the dollar price of the bond is the most sensible metric. As we only interest ourselves in performing credit we ignore this from now on.)
Here is our development: we write down the valuation formula for a bond in terms of survival probability and riskfree discounting; compare and contrast with Z-spread and carefully explain the `par/non-par problem'; define par-adjusted spread; give examples of parametric fitting of the survival curve; move to fitting multiple curves; explain how to calculate carry and rolldown and relative value properly.
But before progressing to the main course we need to munch through one particular nettle: what is the riskfree rate?

\subsection{Riskfree rate}
\label{sec:rfrate}

A good part of the matter's complexity predates the whole LIBOR/OIS situation by decades and is rooted in whether the investor is `real-money' or `levered'. 

The vast majority of bond investing is done by real-money accounts such as pension funds, insurance companies and sovereign wealth funds. For them, the relevance of a spread measure is simply to quantify the excess yield over a Treasury bond of the same maturity (we are USD-focused; in EUR we would use the German government yield). To hedge interest rate risk, they short Treasuries or the futures. In that case the appropriate riskfree rate is simply the Treasury yield, and indeed Duffie~\& Singleton say the same in \cite[\S7.2]{DuffieSingleton03}. Aside from simplicity, this definition confers another advantage: spreads will necessarily be positive. 

As soon as we move to levered investing the position becomes more complicated. The correct discounting rate is OIS and depends on the collaterisation of the borrower. However, even this is not fully correct, because the economics may be more complicated: a hedge fund will typically have a chunk of cash to invest and also a leverage facility, which it can use on demand. When operating without leverage the riskfree rate is the same as for real-money firms, but when the leverage facility is used, the funding rate and collateralisation come into play. To us it seems that the best way of assessing bond spread is to use Treasury as the riskfree rate, and then when a firm needs to borrow money the economics of the trade include the borrowing costs which have their own idiosyncrasies and dynamics.

That said, one can use the swap rates and derive spreads to the swap curve, as is often done, and end up with (usually) a slightly lower answer than the spread to Tsy. This is unlikely to cause practical difficulties, but there is one thing to watch: the spread for the highest-grade issuers may not be positive, as they may be a better credit risk than the banks. If, as here, spreads are required to be positive, then these issuers will always appear expensive.

\subsection{Valuation formulae}
\label{sec:val}

We write $B(T)$ for the riskfree discounting curve and $Q(T)$ for the survival curve. The PV of a survival-contingent payment occurring at some future date $T$ is $B(T)Q(T)$. This allows the coupon stream to be valued as a sum
\[
\sum_j c_j B(T_j) Q(T_j).
\]
If, as previously mentioned, we approximate this as a continuous payment stream, then it is $c\Pi(T)$ where $\Pi$ is known as the RPV01 (risky PV01), where
\[
\Pi(T) = \int_0^T B(t) Q(t) \, dt.
\]
The principal repayment is simply $B(T)Q(T)$---but only if there is no possibly of recovery. In reality if default occurs we can present a claim for the principal amount and expect some proportion $\recov$ (the recovery rate) to be honoured.

To value this extra amount, we divide the time frame $[0,T]$ into slices and note that the probability of default in precisely the interval $[t,t+dt]$ is $-dQ(t)$. Multiplying by $B(t)$ and integrating gives $\recov\,\Xi$, where 
\[
\Xi(T) = -\int_0^T B(t) \, dQ(t).
\]
Both $\Pi,\Xi$ are bilinear forms in\footnote{Paying homage to a well-known British DIY chain.} B\&Q.
Adding the parts gives the model price of the bond:
\begin{equation}
\Phat/100 = c\, \Pi(T) + B(T)Q(T) + \recov\, \Xi(T) .
\label{eq:bp}
\end{equation}
This formula, and the reasoning behind it, emphasises an important distinction between the principal and the coupon stream: the former is partly recoverable, the latter not. Hence the coupon stream is a riskier claim on the firm than the principal repayment, and this is the root of the par/non-par problem. Incidentally we said when discussing \cite{Duffie97} that it makes no sense for the valuation formula to be nonlinear in the LGD: clearly, (\ref{eq:bp}) is linear in $\recov$.

The remainder of this section is devoted to the valuation of credit default swaps.
Following the CDS Big Bang in early 2009 all CDSs trade with a fixed coupon $c$ (100bp or 500bp), so they are not par instruments. When quoted upfront (symbol $u$), there is no difficulty with the valuation and we simply have 
\begin{equation}
u_n + c\Pi_n = (1-\recov) \Xi_n  .
\label{eq:cdsval}
\end{equation}
Names that are quoted with a so-called traded spread are dealt with via an extra step. Loosely the upfront payment must be the product of the difference between the traded spread $\widetilde{s}$ and the coupon, multiplied by some sort of RPV01 $\widetilde{\Pi}$.
To avoid the problem of needing the whole curve to work out the upfront payment, a flat hazard rate curve assumption is used, i.e.\ the same hazard rate $\widetilde{s}/(1-\recov)$ is applied to the zero-rates of the riskfree curve. It might seem that somehow there is a discrepancy between this and the reality of the par curve not being flat, but that is not so: the traded spread differs from the par spread, and also depends on $c$. Thus in the rare case of the same CDS being quoted with two different coupons (100bp and 500bp) there will be two different traded spreads: notice in that case alone we have a market-implied RPV01, by taking the difference of the upfronts and dividing by 400bp.
Practitioners would do well to note that the Bloomberg CDSW screen calculates a `price' $P_\textrm{cds}$ as the price of an FRN with coupon $c$ and spread $\tilde{s}$: that is to say, $P_\textrm{cds}/100 = 1+(c-\widetilde{s})\widetilde{\Pi}$. This quantity is, in our notation, just $-u$, and it can be obtained programmatically via the Bloomberg API.

\subsection{Zero-coupon bonds}

This section might seem pointless, as surely a zero-coupon bond has the same pricing formula, only simpler because $c=0$?
In fact, the position is a good deal more complicated because \emph{accreting bonds}, i.e.\ those not issued at or near par, have different bankruptcy treatment. Put succinctly: the bankruptcy claim is less than 100.

Suppose the bond is issued at a price $P_{00}$. The difference between this and 100 is called the OID (original issue discount). Referring to a legal source\footnote{See {\tt restructuring.weil.com/claims}, ``Original Issue Discount Claims Arising From Fair Market Value Debt Exchanges: Fair Game for Disallowance?'', by Jessica Liou, 02-Oct-13, accessed 12-Aug-23}, which we quote verbatim:

``Despite the OID being paid only when the debt matures, it is, in essence, interest and, from an accounting and tax perspective, OID is typically amortised over the life of the debt and treated as interest in addition to any periodic or regular interest under the debt instrument.
[\ldots] Under section 502(b)(2) of the Bankruptcy Code, a claim for interest that has yet to accrue as of the petition date is generally considered a disallowable claim for unmatured interest.  Section 502(b)(2)'s legislative history unambiguously includes within the meaning of unmatured interest `prepaid interest that represents an original discounting of the claim \ldots that would not have been earned on the date of bankruptcy'.  Thus, courts considering whether to disallow claims for OID agree that the Bankruptcy Code clearly requires them to disallow claims for unamortised OID.''

In other words there is a big difference between buying a low-coupon bond at 70 (which was issued some while ago at par but has since sold off) and a zero-coupon with an OID of 30. In the former, the bankruptcy claim is 100: the fact that it was bought at 70 is an irrelevance, and just rewards the investor for the foresight of not having bought it earlier at a higher price. The latter's claim is somewhere between 70 and 100 depending on the date.

Now for the equations.
Define the accretion factor to be
\[
A(t) = \frac{P_{00}}{100} + \frac{t-T_{00}}{T-T_{00}} \left( 1 -\frac{P_{00}}{100} \right),
\]
which rises linearly from $P_{00}/100$ at the issue date $T_{00}$ to 1 at the maturity date $T$. It is this function that defines the bankruptcy claim; incidentally, it is this function that specifies what the virtual income of the bond is for tax purposes.
Accordingly, the bond pricing equation is the same as (\ref{eq:bp}) with $c=0$, but with the alteration
\begin{equation}
\Xi(T) = -\int_0^T A(t) B(t) \, dQ(t).
\end{equation}

It is worth noting that such bonds are not in general deliverable into CDS, as they would `unfairly' have a lower dollar price than the regular bonds. This should be remembered when attempting the cash-CDS basis trade.

\subsection{The problem with Z-spread}
\label{eq:zspd}

We return to vanilla bonds and their associated spread measures.
Compare (\ref{eq:bp}) with the simple-minded price-yield relationship for a vanilla bond, which does not recognise the concept of recovery:
\begin{equation}
P/100 = c \, \frac{1-(1+y/m)^{-mT}}{y} + (1+y/m)^{-mT} ,
\label{eq:y2p}
\end{equation}
with $m$ the compounding frequency.
This comes from the more general equation for the internal rate of return (IRR) of a set of cashflows,
\begin{equation}
P/100 = \sum_j C_j (1+y/m)^{-mT_j},
\label{eq:irr}
\end{equation}
in the specific case where the coupon amounts are $c/m$, plus the principal at time $T$, and summing over the coupons as a geometric progression. (This is exact for $mT$ an integer, and we use it regardless.)
The Z-spread $s_Z$ is obtained from (\ref{eq:irr}) by replacing $y$ with the sum of two parts, the riskfree zero rate\footnote{The value $z=z(T)$ such that  $B(T)=(1+z/m)^{-mT}$.} and a constant $s_Z$, whose value is uniquely inferred from $P$.

The similarity between (\ref{eq:bp}) and (\ref{eq:y2p}) is that (\ref{eq:y2p}) is a special case of (\ref{eq:bp}) when $B$, $Q$ are both exponential functions.
The difference is that (\ref{eq:y2p}) treats the coupons and principal as essentially the same, but the market clearly does not, as we now explain.

If two bonds of the same maturity but different coupon trade at the same yield, then this means that the discounting mechanism is the same for the principal as for the coupon, i.e.\ just using a risky discount factor, and so the market is pricing zero recovery. Conversely if we consider that recovery will be zero then the two bonds should trade at the same yield: the downside of buying each is the same, and so the upside, as measured by the yield, should be too. We can formalise this by saying that if the high-coupon bond trades at a higher yield then we should buy it and short an equal cash value of low-coupon bond. This will guarantee a profit, if either the issuer survives or else defaults with zero recovery; but if the recovery is intermediate there can be a serious loss, e.g.\ if the bonds are trading at 130 and 85 when the trade is put on, and a little while later default occurs with recovery 90\%, then both legs of the trade lose money.

\begin{figure}[!h]
\begin{center} \begin{tabular}{ll}
(i) & \scalebox{0.8}{\includegraphics{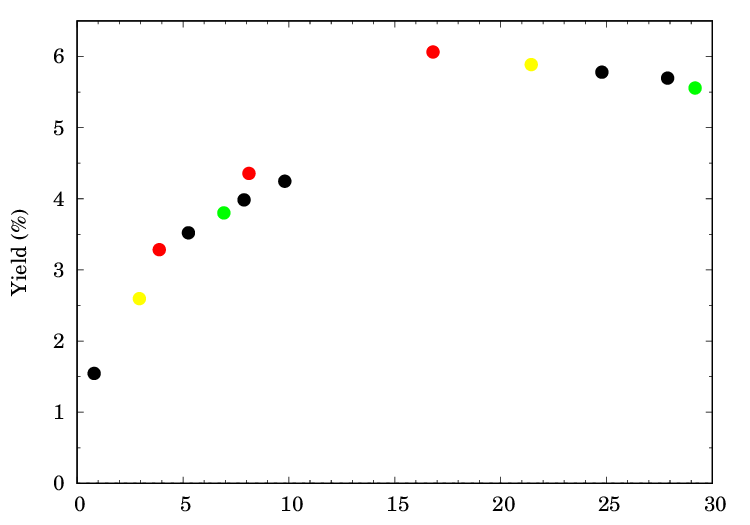}} \\
(ii) & \scalebox{0.8}{\includegraphics{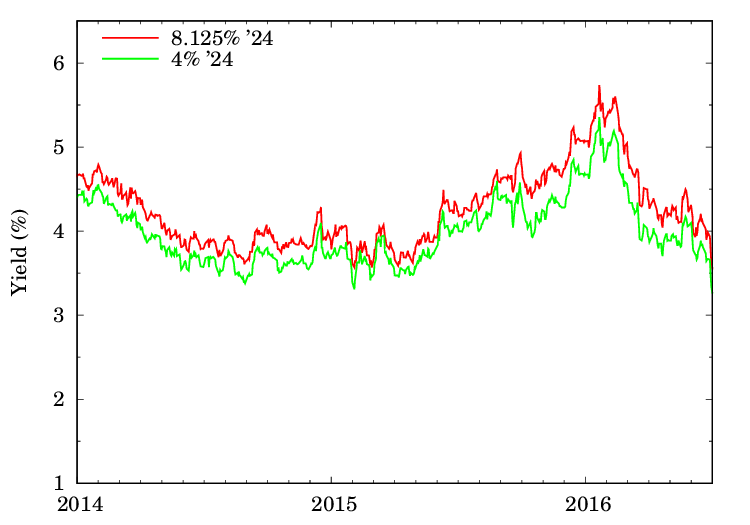}} 
\end{tabular}
\end{center}
\mycaption{(i) COLOM \$ bonds on 08-Apr-16: yield vs tenor. Red, yellow, black, green respectively denote bonds trading above 120, around 110, around 100, and around 90 dollar price.
(ii) Comparison of 8.125\% '24 vs 4\% '24 over time shows a consistent yield (spread) difference.}
\label{fig:colom}
\end{figure}

But in general the market does not price bonds this way, and so the high-coupon bond trades at a higher yield. See for example Figure~\ref{fig:colom}, illustrating a common effect. 
In general the market thinks as follows: if spreads/yields grind tighter then the dollar price will rise further and become even less attractive to buy, while if something bad happens to the issuer then the high-dollar price bonds stand to lose more. In other words there is some negative convexity embedded in the bond, despite the fact that it is not callable.

For more precise analysis we need to find bonds of widely different coupon but almost identical maturity, but that situation is uncommon. As bonds are nearly always issued at par, it only happens when the bonds were issued a long time apart, with maturity dates that happen by chance to (almost) coincide, and when yields have moved a long way in the interim.

\begin{figure}
\centering
\begin{tabular}{lrrr}
\hline
Coupon(\%) & 4 & 8.125 \\
Tenor(y) & 7.88 & 8.11 \\
Price & 100.10 & 125.50 \\
Yield(\%) & 3.98 & 4.36 \\
\hline 
$\recov$ & $\Delta P$ & $\Delta P$ & $\lambda$ \\
\hline 
0\% & $-1.52$ & $+1.52$ & 0.0281 \\ 
20\% & $-1.18$ & $+1.18$ & 0.0346 \\
40\% & $-0.65$ & $+0.65$ & 0.0451 \\
50\% & $-0.27$ & $+0.27$ & 0.0531 \\
55\% & $-0.03$ & $0.03$ & 0.0582  \\
60\% & $+0.26$ & $-0.26$ &  0.0644 \\
70\% & $+1.00$ & $-1.00$ &  0.0819  \\
\hline
\end{tabular}
\mycaption{
Two COLOM '24 bonds as of 08-Apr-16, valued with different recovery assumptions. $\Delta P$ is the model price minus the market price (hence, positive means bond appears cheap).
}
\label{tab:colom}
\end{figure}

A good example is Colombia, which in 2004 issued 8.125\% USD bonds due 21-May-24, and in 2014 issued 4\% bonds due 26-Feb-24. The yield difference in Figure~\ref{fig:colom}(ii) is reasonably static over time, and cannot simply be ascribed to liquidity\footnote{It is a convenient fact that the author traded both of them over the period shown.}.
We now value these bonds as of 08-Apr-16 using a flat hazard rate $Q(T)=e^{-\lambda T}$, with different assumptions about recovery.
In each case the hazard rate $\lambda$ is adjusted so as to make the total pricing error $\Delta P_1+\Delta P_2$ zero,  where $\Delta P_j= \Phat_j-P_j$.
Table~\ref{tab:colom} shows the results\footnote{USD zero-rates for maturity 0,1,2,\ldots,10y are taken as (semiannual, in \%) 0.65, 0.74, 0.85, 0.94, 1.05, 1.15, 1.26, 1.37, 1.46, 1.55, 1.55 with linear interpolation in between. Time increment (see \S\ref{sec:comp}) is 0.5y.}.
For $\recov=0$ the model price is too low for the 4\% bond and too high for the 8.125\% bond, as expected. As $\recov$ is raised the error reduces until at $\recov=55.5\%$ both match, and this is the implied recovery.
If we think that Colombia recovery should be lower than this figure, then we buy the 8.125\% and enjoy the higher yield while it lasts; if higher, then we should buy the 4\%. In the limit of $\recov\to100\%$, which seems silly but could in principle happen in a `technical' default, the 4\% become risk-free, so these are obviously the ones to buy. We have, incidentally, ruled out the possibility of a sovereign selectively defaulting on some of its bonds: with corporates issued under NY or UK law this cannot happen but with EM sovereigns ``one can never say never''.

When building curves we should not, therefore, simply use yield or yield-related spread measures such as Z- or I-spread. Although it is too fiddly to attempt to infer the market-implied recovery when we have many bonds of different maturity and dollar price, we should make a better attempt than simply assuming the recovery to be zero. Otherwise, high dollar price bonds always look cheap, when in fact much of the cheapness is illusory. In the case of the two COLOM~'24 bonds, if we make the routine assumption of 40\% recovery, we will have the 8.125\% modelled about a point cheap to the 4\% bonds---still a gap, but only a little wider than the bid-offer spread (which would typically be around 0.5--0.75\,pts), and so the transaction cost of switching from the 4\% to the 8.125\% would scarcely be worth the bother, particularly when we remember that high dollar price bonds tend to be more difficult to trade. But if we assume $\recov=0$ then there appear to be almost 3\,pts of value in switching.

\subsection{Comment on other spread measures}

We have already discussed Z-spread. Other measures are available but all have their own problems; we briefly discuss them now.
One way in which spread definitions differ is in their assumptions about the RPV01.

The \emph{benchmark spread} or spread to Treasury (SoT) is the most rudimentary of the yield-based measures, and is almost uinversal in USD markets. It is simply the yield difference between the bond in question and the benchmark Tsy bond which is \emph{not} maturity-matched. Thus bonds of maturity 7--15y are all quoted off the 10y Tsy. This is helpful for pricing on the day of the trade but it is a useless construct for analytical work. As a bond's maturity moves down to around 7y, it jumps to being quoted off the 5y Tsy, which typically has lower yield than the 10y, and so the SoT suddenly jumps up even if the bond price has not moved!

The \emph{asset-swap spread} of a fixed-rate bond can be thought of in a couple of equivalent ways. One is that I can hand over the coupons and an upfront amount $1-P/100$ (if this is negative, I receive money) to a swap counterparty in return for LIBOR plus a spread $s_A$, which is the asset-swap spread. Consequently, with $R$ denoting the par swap rate, and $\Pi^\circ_B$, $\Pi^\circ_F$ the fixed and floating swap PV01s:
\begin{equation}
s_A = \frac{\displaystyle 1- \frac{P}{100} + (c-R) \Pi^\circ_B}{\Pi^\circ_F} .
\label{eq:asw}
\end{equation}
This is linear in the bond price, because the numerator of the above equation is the riskiness of the bond in price terms---it increases if $P$ is lower or $c$ is higher, and is zero if the bond trades flat to LIBOR---but the denominator is the swap PV01, which has nothing to do with the bond's credit quality. 

Finally the par CDS spread is the value of the running spread that makes a CDS contract value to par. It is similar to the Z-spread in the sense that if the CDS trades off the same survival curve as the bond (i.e.\ there is no basis between the two markets) then, as the survival probability declines, and with it the bond price, the par spread increases and is a convex function of the bond price. The difference is that the CDS spread hits $\infty$ when the bond price hits recovery; so typically it exceeds the Z-spread.

\begin{figure}
\begin{center}
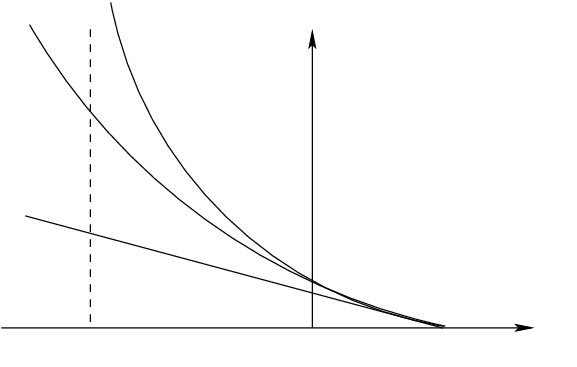
\end {center}
\caption{Asset-swap spread, Z-spread and CDS spread (schematically) vs bond price. All are zero when the bond price is $P_{RF}$, the value of a bond with the same payment schedule but no credit risk.
Note the different asymptotes as $P\to0$.}
\label{fig:ch4.zaswcds}
\end{figure}

\subsection{Par-adjusted spread}

One might suppose, given that valuation of bonds can be done without reference to any kind of spread measure (one needs only $B$, $Q$ and a recovery assumption), that we can scrap the whole idea of spread altogether. The problem with that is that everybody thinks about in spread terms, and we can scarcely abandon the standard tool of plotting spread vs maturity (or duration). We have said that the Z- and I-spreads are not right, but then what is the correct idea? We pursue this now.

From (\ref{eq:bp}) we have, using the par CDS spread $s(T) = (1-\recov) \Xi(T)/\Pi(T)$:
\begin{equation}
\widehat{P}/100 - 1 = \biggr[ c - \underbrace{ \left( \frac{1-B(T)Q(T)-\Xi(T)}{\Pi(T)} \right) }_{\rhat(T)}   - s(T) \biggr] \Pi(T) .
\end{equation}
The LHS is the deviation from par of the (model) bond price.
On the RHS the term in square brackets is understood as follows: the coupon minus a sort of riskfree rate, to be explained presently, minus the par CDS spread. This is intuitively reasonable because if the bond is trading at par then its asset-swap spread (coupon minus swap rate) should equal the par CDS spread, to avoid arbitrage.
We now explain the second term, which we will write as $\rhat(T)$, in more detail. For a start we can neatly recast it as the \emph{parity equation}
\begin{equation}
B(T)Q(T) + \Xi(T) + \rhat(T)\Pi(T) = 1.
\label{eq:par}
\end{equation}
Next, the numerator of $\rhat(T)$, which is $1-B(T)Q(T)-\Xi(T)$, can be written (integration by parts)
\[
1 - B(T)Q(T)  + \int_0^T B(t) \, dQ(t) = -\int_0^T Q(t) \, dB(t),
\]
and recalling the definition of the riskfree instantaneous forward rate $f(t) = -B'(t)/B(t)$, we find
\[
\rhat(T) = \frac{\int_0^T f(t) B(t)Q(t) \, dt}{\int_0^T B(t)Q(t) \, dt}, 
\]
the average of instantaneous riskfree forward rates weighted by the risky discount factor. In the case of a `flat riskfree curve',  $f$ is constant, and as $Q$ then divides out, $\rhat$ is just equal to $f$. When $Q\equiv 1$, $\rhat$ is just the riskfree par rate.
Now using the actual bond price $P$ we can write
\begin{equation}
P/100 - 1 = \big( c - \rhat(T)  - \sbar \big) \Pi(T) ,
\label{eq:parspdbond}
\end{equation}
which defines $\sbar$, and we call it the \emph{par-adjusted spread} of the bond. Note that $\sbar-s(T)$ is the basis in spread terms between cash and CDS\footnote{The so-called negative basis trade is where the cash bond trades cheap to CDS: in our notation, $\sbar-s>0$ then.}.

The par-adjusted spread of a CDS is just the par spread. If the CDS is quoted upfront $u$ plus a fixed running spread of $c$, then the par-adjusted spread is
\[
\sbar=c+u/\Pi,
\]
which is the coupon plus the upfront converted into a running spread. If instead the CDS is quoted as a spread $\widetilde{s}$ then, as we know from previous discussion, we convert it to an upfront using the standard mechanism.
Then, analogously with (\ref{eq:parspdbond}), we have
\[
P_\textrm{cds}/100 -1 = ( c - \sbar ) \Pi(T) ,
\]
the main difference being the absence of the $\rhat$ term, which refers to the funding of the bond, but is not needed for a synthetic position. A different way of expressing this is the conversion of the traded spread $\widetilde{s}$ into the par spread:
\[
(\sbar - c)\Pi = (\widetilde{s}-c)\widetilde{\Pi}
\]
with $\widetilde{\Pi}$ the ISDA `flat-hazard-curve' RPV01.

We are now in a position to write down the expression for the deviation in price terms between the (par-adjusted) spread of a bond $\sbar$  or CDS and a curve giving $s(T),\Pi(T)$:
\begin{equation}
\Delta P/100 = 
\big(\sbar-s(T)\big) \Pi(T) = 1-P/100 + \big( c-\rhat(T)-s(T) \big) \Pi(T)
\label{eq:resid}
\end{equation}
with $\rhat$ omitted for a CDS. Roughly, in fitting to price data, we can simply choose to minimise the total squared residual $\sum_j(\Delta P_j)^2$, but we shall talk more about this shortly in \S\ref{sec:fit}.

\subsection{Computational matters and CDS stripping}
\label{sec:comp}

The expressions for $\Pi$, $\Xi$ involve integrals, which in practice we approximate using the trapezium rule: in detail,
\begin{eqnarray}
\Pi(t_N) &\approx& \sum_{n=1}^N \frac{B(t_{n-1})Q(t_{n-1})+B(t_n)Q(t_n)}{2}  (t_n-t_{n-1}) \nonumber \\
\Xi(t_N) &\approx& \sum_{n=1}^N \frac{B(t_{n-1})+B(t_n)}{2} \big(Q(t_{n-1})-Q(t_n)\big) \label{eq:compute1} \\
\rhat(t_N)\Pi(t_N) &\approx& \sum_{n=1}^N \big(B(t_{n-1})-B(t_n)\big) \frac{Q(t_{n-1})+Q(t_n)}{2}  \nonumber .
\end{eqnarray}
These approximations exactly satisfy the parity equation (\ref{eq:par}), which incidentally uses only the second and third of them. For most of the work in this paper the interval spacing is 1y.

A minor difficulty occurs when the bond is perpetual, or of maturity longer than the end-point of our computational grid, which we assume to be set up to compute as far as a maximum maturity $T^\star$ (something like 25--30y): there are always a few awkward bonds in the universe\footnote{For example the PETBRA 100y bond issued a few years ago, as well as a variety of long-dated sovereigns.} whose maturity exceeds it.  
By assuming that riskfree forward rate and forward hazard rates are flat for maturity $>T^\star$ we have
\begin{eqnarray*}
\Pi(T) &\approx& \Pi(T^\star) + \Delta \Pi \\ 
\Xi(T) &\approx& \Xi(T^\star) + \lambda \, \Delta \Pi  \\
\Delta \Pi &:=& B(T^\star)Q(T^\star) \frac{1-e^{-(r+\lambda)(T-T^\star)}}{r+\lambda} .
\end{eqnarray*}

Different considerations apply when one is dealing with CDSs only, we have a full curve of quotes (or at least the main points such as 1,2,3,5,7,10y), and wish to `strip the curve'. All the payment dates are standardised to 20th March, June, September and December, and one usually has some sort of full curve. It is then natural to ask for the survival curve to be calculated on this grid of dates, or more practically on the subset corresponding to integer maturities. In other words, if we are in April 2024 then the maturity dates are 20-Jun-25, 20-Jun-26 and so on. Only in the case of names that are expected to default very soon are we likely to see quotes for 20-Jun-24, 20-Sep-24 etc.

It is common practice to use a slightly different type of interpolation than what is shown above. That is to say, in between points we assume the forward hazard rate $\lambda$ to be piecewise constant, which corresponds to $\ln Q(T)$ being linear; similarly $B(T)$ is given the piecewise-exponential treatment too. The necessary integrations can then be done exactly.

In detail, between two coupon dates $t_{j-1}$, $t_j$, the contribution to the RPV01, i.e.\ the payment at the end of the interval, is one of the following two formulae dependending on the contract specification. If accrued interest is not payable when default occurs partway through the interval, we have the first expression; otherwise, the second:
\begin{eqnarray}
\Pi_{t_{j-1},t_j} &=& B(t_j) Q(t_j) \alpha(t_j - t_{j-1}) \qquad \mbox{(without accrued)} \nonumber \\
\Pi_{t_{j-1},t_j} &=& B(t_j) \int_{t_{j-1}}^{t_j}  Q(t) \,\alpha\,dt \qquad \mbox{(with accrued)}   \\
&=& B(t_j) Q(t_{j-1})  \mathcal{E}(\lambda(t_j-t_{j-1})) \, \alpha(t_j - t_{j-1})  \nonumber
\end{eqnarray}
Here $\mathcal{E}(x) \equiv (1-e^{-x})/x$.
The second specification is more usual.
Notice that $\mathcal{E}(x) = \half(1+e^{-x}) + O(x^2)$ for small $x$, and so the last expression is roughly $B(t_j)(Q(t_{j-1})+Q(t_j))/2$, which is what motivated the approximation for $\Pi$ in (\ref{eq:compute1}).

The symbol $\alpha$ denotes the accrual or coverage factor, i.e.\ the year fraction represented by the time interval. For CDS this is Act/360, so if the $t$'s are in days then $\alpha=\frac{1}{360}$. This is the only concession we make to day-count conventions\footnote{For the pricing of interest rate swaps where notionals are very much higher, a day's interest makes a difference and day-count and holiday calendars are important. For these purposes they are not. However, the distinction between Act/360 and Act/Act on a 500bp coupon is worth including given it requires so little effort.}, as we approximate all other conventions such as 30/360, Act/365, Act/Act as Act/365.25, i.e.\ $\alpha=\frac{1}{365.25}$.

Now in practice if the quotes are given in yearly steps then we make $B,Q$ exponential in yearly pieces and sum four (or more generally $m$, but CDS are always quarterly) of these expressions together. This is just a geometric progression, and we have, where we have now changed our notation so that the interval is a year long and denoted $[t_{n-1},t_n]$:
\[
\Pi(t_{n-1},t_n) = B(t_{n-1}) Q(t_{n-1}) \frac{ e^{-r(t_n-t_{n-1})/m} \mathcal{E}(\lambda(t_{n}-t_{n-1})/m) }{ \mathcal{E}((r+\lambda)(t_{n}-t_{n-1})/m) } 
 \mathcal{E}((r+\lambda)(t_{n}-t_{n-1}))  \alpha (t_n - t_{n-1})
\]
(where $B(t_n)=B(t_{n-1})e^{-r(t_n-t_{n-1})}$);  as $m\to\infty$ we arrive at the simpler expression
\begin{equation}
\Pi(t_{n-1},t_n) \approx B(t_{n-1}) Q(t_{n-1}) 
 \mathcal{E}(t_{n-1},t_n) \, \alpha 
\end{equation}
with $\mathcal{E}(t_{n-1},t_n) = (1-e^{-(r+\lambda)(t_n-t_{n-1})})/(r+\lambda)$.
The payout leg presents no difficulties and we have simply
\begin{equation}
\Xi(t_{n-1},t_n) = B(t_{n-1}) Q(t_{n-1}) \mathcal{E}(t_{n-1},t_n) \, \lambda .
\end{equation}

For ease of exposition we use these last two expressions and now show how to strip the curve, i.e.\ obtain the $Q$'s recursively. Writing for short $\Pi_n = \Pi(t_n)$, etc., we reexpress (\ref{eq:cdsval}) as 
\begin{eqnarray*}
\Pi_n &=& \Pi_{n-1} + B_{n-1}Q_{n-1} \alpha \mathcal{E}_n \\
\Xi_n &=& \Xi_{n-1} + B_{n-1}Q_{n-1} \lambda_n \mathcal{E}_n \\
\mathcal{E}_n & := & \frac{1-e^{-(r_n+\lambda_n)(t_n-t_{n-1})}}{r_n+\lambda_n}  
\end{eqnarray*}
with $r_n = -(t_n-t_{n-1})\inv \ln(B_n/B_{n-1})$.
The valuation equation (\ref{eq:cdsval}) can be recast as
\begin{equation}
\frac{u_n + c\Pi_{n-1} - (1-\recov)\Xi_{n-1}}{B_{n-1}Q_{n-1}} = \big((1-\recov)\lambda_n - \alpha c)\big) \mathcal{E}_n
\label{eq:strip2}
\end{equation}
in which the LHS is known, as the computed terms all bear subscript $_{n-1}$. Now $\lambda_n$ appears directly and also through $\mathcal{E}_n$, and a practical solution to this problem is the following recursion: (i) start with some estimate of $\lambda_n$, such as zero, or if $n>1$ we can use $\lambda_{n-1}$; (ii) use $\lambda_n$ to find $\mathcal{E}_n$; (iii) with $\mathcal{E}_n$ fixed, solve for $\lambda_n$ using (\ref{eq:strip2}), which is linear; go back to (ii) and repeat until it converges. It is easily seen that the LHS must lie in the range specified as follows: the lower bound is when $\lambda_n=0$ and given by $-c(1-e^{-r(t_n-t_{n-1})})/r$, the upper when $\lambda_n=\infty$ and given by $1-\recov$.
Before attempting to solve for $\lambda_n$ it is sensible to check that these arbitrage bounds are not being violated.

\subsection{Fitting}
\label{sec:fit}

Suppose that we have a model for $Q(T)$, parametrised somehow: this will automatically imply $s(T),\Pi(T)$ and thence a model bond price.  We want to minimise the fitting error with respect to the parameters. This section discusses some minor matters relating to that problem.

First, we will typically want to weight the errors by issue size $N_j$ say\footnote{Technically, the amount outstanding in million USD. We assume this to be \$1Bn for liquid CDS.}.
Next, should we use price or par-adjusted spread: i.e.~should there be a duration weighting on the residuals (\ref{eq:resid})? We can write down three possibilities:
\begin{itemize}
\item[(i)]
$\sum N_j (\Delta P_j)^2$;
\item[(ii)]
$\sum N_j (\Delta s_j)^2$;
\item[(iii)]
$\sum N_j (\Delta s_j)^2 \Pi_j = \sum N_j (\Delta P_j)^2 /\Pi_j$.
\end{itemize}
The third is intermediate, in the sense of weighting long-duration bonds more than (ii) but less than (i).
It is worth noting that both (ii) and (iii) can get tripped up by stale prices on very short-dated bonds, e.g.\ if the market price is allegedly 100.5 but the bond has only one day to maturity: then in going from (i) to (iii) the pricing error has suddenly been multiplied by 365, and to (ii) a second factor of 365. One way to resolve this is simply to ignore bonds of maturity less than 1y or so, or in (iii) to use the RHS (squared pricing error, divided by RPV01) and floor the RPV01 at 1y. Arguably (i) pays too much attention to longer-dated bonds. However, experience suggests that, as a result of there being a parameter that directly influences short-dated bond pricing (this being parameter $a$), short-dated bonds are not ignored. It is also worth noting (see Figure~\ref{fig:miningb+c}) that there are more short-dated bonds than long-dated ones in the universe, which provides an extra weight to the subset with shortest maturity, and the author's preference is (i)$>$(iii)$\gg$(ii).
A final matter is to improve over simple least squares: use a penalty that increases less rapidly than quadratically for large deviations, e.g.\ $\sqrt{1+(\Delta P)^2}-1$. A fuller discussion of the general principle is in \cite[\S15.7]{NRC}. 

Returning to the two COLOM bonds, when $\recov=53.5\%$ we price both exactly, using the same hazard rate, and the par-adjusted spreads are both 260\,bp. For these assumptions neither bond is `better' than the other, as their maturities are virtually identical and they have the same par-adjusted spread. But if we use $\recov=0$ we need different hazard rates to price the bonds: for the 4\%, we have $\lambda=0.0256$ and $\sbar=258$\,bp, and for the 8.125\%, we have $\lambda=0.0296$ and $\sbar=298$\,bp. The difference in par-adjusted spread (40\,bp) is similar to the yield difference. In conclusion, the par-adjusted spread measure takes recovery and deviation from par into account.

\subsection{Hazard curve model: Single-name case}

In general a constant forward hazard rate will not be sufficient to capture the term structure accurately, so we need to fit a curve, and are looking for a sensible parametrisation.
We first deal with the case where we have one name, and enough bonds to make it sensible to fit a curve. If this is not so then we need to fit multiple names and maturities in the sector and, in effect, interpolate based on internal or external credit rating. This is covered by the next subsection. After that we can always adjust one parameter so as to exactly fit the model curve to a given bond.

A model that seems to offer the right amount of flexibility, while giving a sensible shape of curve, is
\begin{equation}
Q(T) =  (1+\gamma T)^{2(b-c)/\gamma} e^{-(a+b-2c)T/(1+\gamma T) -bT} ,
\qquad
-\frac{Q'(T)}{Q(T)} = \frac{a+2c \gamma T + b(\gamma T)^2}{(1+\gamma T)^2}
\qquad
\label{eq:Q1}
\end{equation}
(the second expression being the instantaneous forward hazard rate, IFHR).
This is an extension of the formula presented in the earlier versions of this paper\footnote{What is called $\gamma$ here was called $c$ in the earlier versions.}. In essence the idea is to write $Q'/Q$ as a polynomial in $u=(1+\gamma T)\inv$. Here it is quadratic but in the previous version it was simply linear, and obtained by writing $c=(a+b)/2$: in that case, the expression for the survival probability reduces to $Q(T)=(1+\gamma T)^{(b-a)/\gamma}e^{-bT}$. More generally this technique of transforming $T\in[0,\infty)$ to $u\in(0,1]$ and using some basis of polynomials on $[0,1]$ is a standard idea in spectral methods \cite{Boyd01} and in the approximation of special functions, e.g.~\cite[\S26.2.16/17]{Abramowitz64} for the cumulative normal distribution. We suggest, however, that polynomials of higher degree than two is likely to result in overfitting.

The interpretation of the parameters is straightforward, as follows. The short-term hazard rate is $a$ and the long-term one is $b$, while $c$ allows the middle of the curve to be altered; the IFHR at $T=\gamma\inv$ is $(a+b+2c)/4$, which is the average of $(a+b)/2$ and $c$. Finally $\gamma$ controls the rate of transition from short to long term. A similarity to the Nelson--Siegel model \cite{NelsonSiegel87} is worth noting in passing.

\begin{figure}[!h]
\hspace{-20mm}\begin{tabular}{ll}
\scalebox{0.75}{\includegraphics{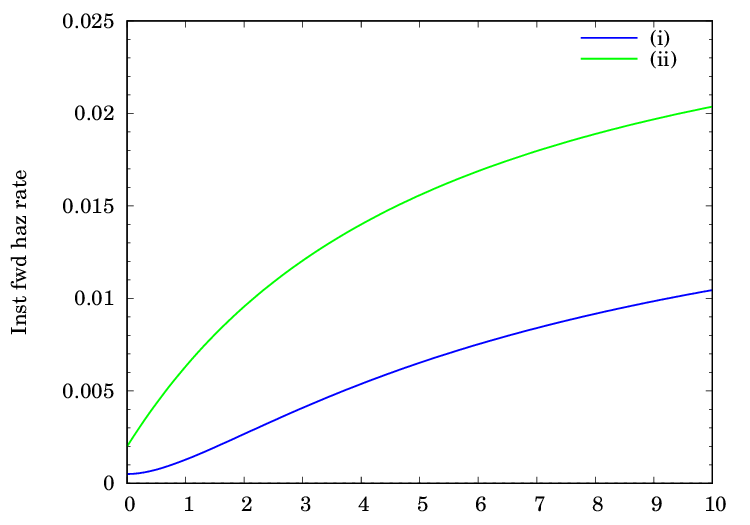}} &
\scalebox{0.75}{\includegraphics{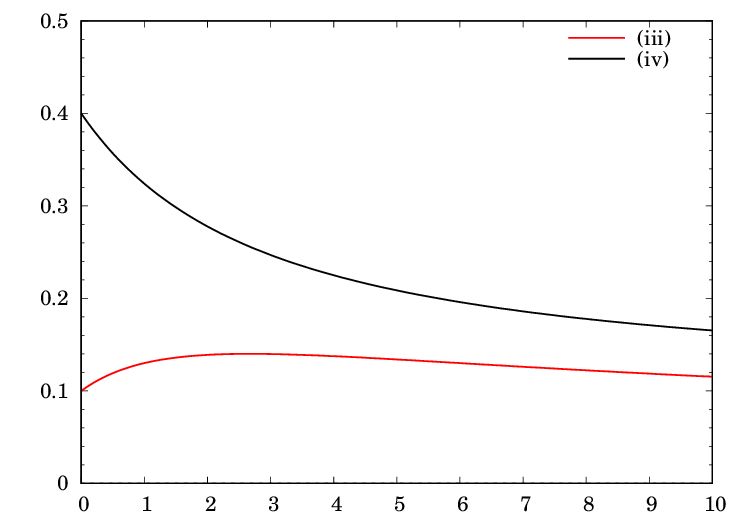}} 
\end{tabular}
\mycaption{Instantaneous forward hazard rate curve $-Q'(T)/Q(T)$ vs $T$ for different parameter sets, showing upward-sloping curves for typical A and BBB credits ((i),(ii) resp.; note the former is inflected);
humped and inverted curves for typical B and CCC credits ((iii),(iv) resp.).
}
\label{fig:ifhr}
\end{figure}

It is necessary (for $Q'<0$, and hence absence of arbitrage) for $a,b,\gamma>0$, but for $c$ it is more complicated.
Various shapes are possible. Note that:
\begin{itemize}
\item
provided $c>-\sqrt{ab}$ the IFHR is positive for all $T>0$, and for avoidance of arbitrage this is the minimum possible $c$; 
\item
the gradient of the IFHR curve at the origin is $2(c-a)\gamma$;
\item
if $c$ does not lie between $a,b$ then the IFHR curve has a stationary point at $T=(c-a)/(c-b)\gamma$.
\end{itemize}
Figure~\ref{fig:ifhr} shows typical examples.
For curves that are essentially upward-sloping ($a<b$) it is unlikely that the front-end gradient will be negative, though it is possible that the curve is deformed upwards at the short end by excessive buying of short-dated CDS protection, as happened in 2008. In general, however, it is likely that we will want $c>a$. The curve will be humped when $c>b$, but for investment-grade curves this is uncommon. When $c<(3a+b)/4$ the curve is inflected, having positive convexity for short maturities and negative for longer ones: otherwise it has negative convexity throughout. The former kind of shape is common in high-grade CDS. 
For curves that are essentially inverted ($a>b$), humped curves are more common. In this case it will be more likely that $c>b$ rather than the more restrictive $c>a$. 
In summary, the suggested protocol is to impose the constraint $c\ge a$; if after fitting it is found that $c=a>b$ then the constraint is altered to $c\ge b$ and the fitting rerun from its current point to see if any improvement can be gleaned.

\begin{figure}[!h]
\begin{center}\begin{tabular}{ll}
(i) & \scalebox{0.8}{\includegraphics{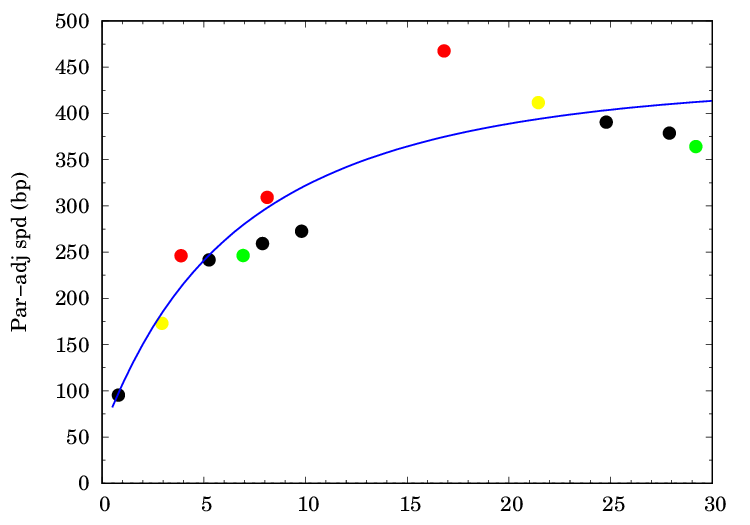}} \\
(ii) & \scalebox{0.8}{\includegraphics{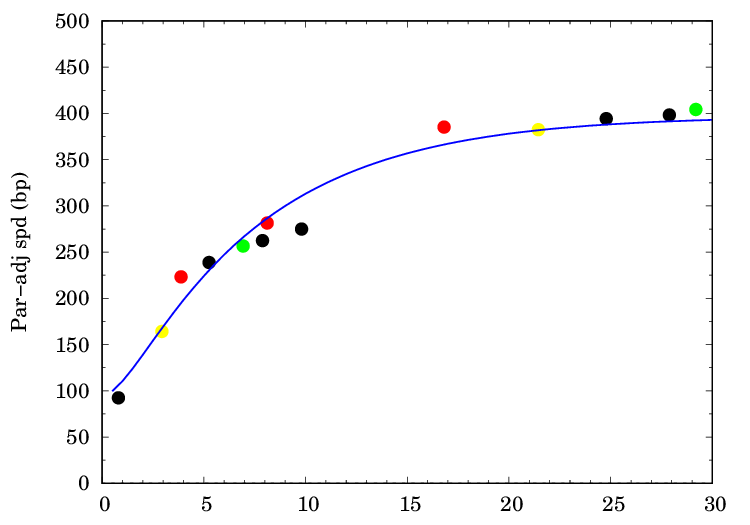}}
\end{tabular}
\end{center}
\mycaption{COLOM \$ bonds on 08-Apr-16: par-adjusted spread and fitted curve. Parameters: (i) $\recov=0$, $a=0.0055$, $b=0.0676$, $c=0.0244$, $\gamma=0.3$;
(ii) $\recov=50\%$, $a=c=0.0190$, $b=0.1718$, $\gamma=0.3$.}
\label{fig:colom2}
\end{figure}


Figure~\ref{fig:colom2} shows the results for the entire set of COLOM bonds, using (i) $\recov=0$ and (ii) $\recov=50\%$. Note how the 10.375\% '33 (145 dollar price, seen on the graph at $T\approx17$y), lies a long way off the curve when $\recov=0$, for the same reason that it does in Figure~\ref{fig:colom}(i), but not when $\recov=50\%$. Indeed, the fit is generally better in (ii), probably because the latter is a more realistic recovery assumption, and hence a better approximation to the way the market works.

\subsection{Hazard curve model: Multiple ratings}

This is the more general form of the model. We now want to fit many ratings at once, and allow the parameters $a,b,c$ to be rating-dependent:
\begin{equation}
Q_j(T) = (1+\gamma T)^{2(b_j-c_j)/\gamma} e^{-(a_j+b_j-2c_j) T/(1+\gamma T) -b_j T}, \quad 
-\frac{Q_j'(T)}{Q_j(T)} =  \frac{a_j+2c_j\gamma T+b_j(\gamma T)^2}{(1+\gamma T)^2} .
\label{eq:Q2}
\end{equation}
It is a useful feature of the way that credit markets seem to work that, roughly, spreads vary in geometric progression across the linear rating scale
\[
\mbox{
AAA=1, AA+=2, AA=3, \ldots, BBB=9, \ldots, CCC=18.
}
\]
This is helpful because when we come to fit to data we cannot have dozens of parameters. This extension of the previous model uses twelve and we suggest that no more be used:

\begin{itemize}
\item
Forward hazard rates $a,b,c$ for rating A (=6)
\item
Forward hazard rates $a,b,c$ for rating BBB (=9)
\item
Forward hazard rates $a,b,c$ for rating BB (=12)
\item
Forward hazard rates $a,b,c$ for rating B (=15)
\item
(Scale parameter $\gamma$, assumed equal across all ratings, which can probably be fixed).
\end{itemize}
This allows us to capture a wide variety of shapes and also, through their time variation, the global movements of all curves up and down, steepening/flattening, and compression (i.e.\ high yield spreads decrease and investment grade increase) or decompression.
In between these four ratings, we use logarithmic interpolation; outside, we use logarithmic extrapolation.

It is a sensible precaution to force a minimum spacing on the parameter set across ratings, in the sense that $a_\textrm{BBB} / a_\textrm{A} > $ some number $>1$, and similarly with $b,c$ and with the other ratings. This ensures that the curves cannot cross, and provides stability to the calibration procedure when there are not many bonds.

\begin{figure}[!h]
\centering
\scalebox{0.8}{\includegraphics{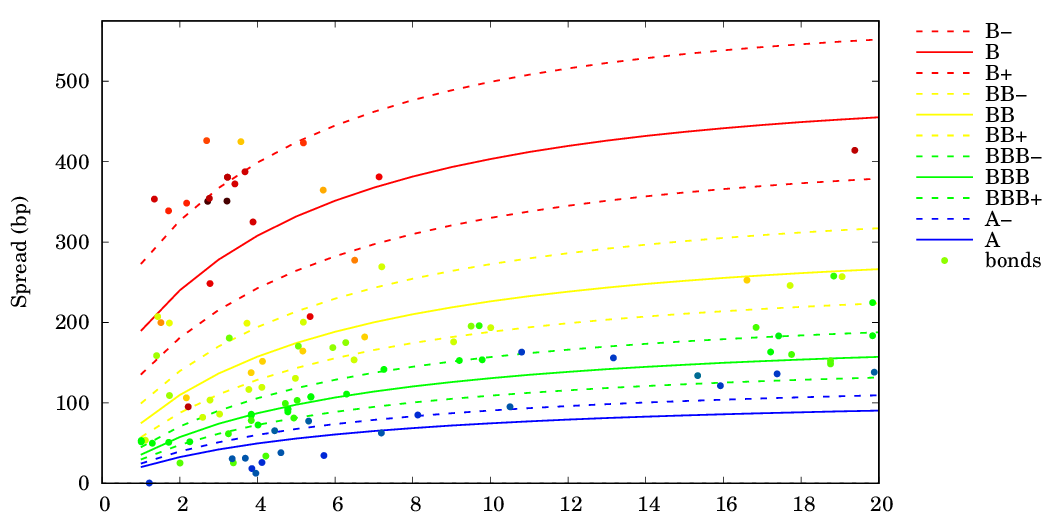}} \\
\mycaption{Fitted spread curves for the mining sector, Jan'18. Bonds shown in dots using same colour scheme as rating curves. Note the dispersion around the fitted curves.}
\label{fig:miningb+c}
\end{figure}

How much dispersion is there when we fit the curves? It is indeed naive to suppose that bonds will line up nicely with their appropriate rating curves, but is the deviation quite small, or does a plot of spread vs maturity, using different colours for different ratings, look more like a swarm of multicoloured bees emerging from a hive? Those with market experience will be more likely to take the latter view, and indeed this is close to reality. The same point was made in relation to risk-neutral calibration of credit migration models, in \cite{Martin21a}. See Figure~\ref{fig:miningb+c} for the mining sector; other sectors are similar.

We have already suggested that the market does not always respect external ratings, as it is providing a constantly-evolving view of credit risk. Some of the dispersion that is found can simply be attributed to external ratings being very out of date. 
A more refined approach is to use the same rating scale for our internal ratings, and then when we are doing the fitting we use those instead. For example, in recent years PEMEX was externally rated BBB$-$ while trading at 400+bp spread; this level, and the credit metrics, were more consistent with BB$-$. By using that as the rating we will typically get a better fit (lower fitting residual) and big issuers that are `clearly misrated' will not distort the fitting procedure. The advantage of using the same rating scale is that where we are agnostic about the rating---and we cannot have an opinion on every issuer under the sun---we can simply use the external one.

\subsection{The need for variable recovery}

We have explained why it is necessary to use a positive recovery rate rather than assuming it to be zero. However, making a simple 40\% recovery assumption, in line with CDS single-name and index pricing, is not ideal either. 
For bonds trading at very low dollar price, a 40\% recovery assumption will make it impossible to fit a curve that agrees with the market price. If used for relative value, the model will show these bonds as cheap irrespective of the credit rating.
To take a specific example, a 1y bond with a high market-implied default probability of $60\%$, assuming $\recov=40\%$, will have a price $P\approx64$, but one can typically find bonds trading lower than that. Arguably models of this sort are no longer useful once a bond becomes distressed, as all bonds will trade at or near recovery, which in turn is completely idiosyncratic, but some improvement on an artificial 40\% seems desirable. 
One solution is to use recovery that declines with rating, and a reasonable construction is to make the recovery $(70-3r)\%$, with $r$ the linear rating. Thereby BBB$-$ has 40\% recovery, AA has 61\%, BB has 34\%, B has 25\%. There is a case for having sovereigns a little higher than this.

\subsection{Relative value, Carry, and Rolldown}

We now turn to problem C as identified at the outset.
Here is a subtle matter worthy of attention.
Suppose a bond trades at a significant spread to its corresponding curve. We can say that this gives rise to extra carry by comparison with bonds that lie on the curve, but also that the bond offers relative value, by being `cheap to the curve'. However, in quantifying the total return we must be careful not to double-count. An obvious way to avoid this is to say that over a period $[0,\Delta t]$ the bond earns extra carry than a bond on the curve, but the relative value component is obtained at the \emph{end} of the period. Loosely this means that the relative value is the spread difference multiplied by the forward duration\footnote{Ambiguous as the forward RPV01 is commonly understood to mean $\Pi(T)-\Pi(\Delta t)$, whereas we mean $\Pi(T-\Delta t)$ here.}, i.e.\ the duration as seen at time $\Delta t$, not as seen today: otherwise we will double-count.

We write $c'=c-\rhat$ for a bond, and $c'=c$ for a CDS, so that the formulae below apply equally to both.
Reminder: $\sbar=$ par-adjusted spread, $\shat=$ model par spread, $\widetilde{s}=$ traded CDS spread.

\begin{itemize}

\item
Credit carry is the change in value, plus accrued coupon, in the event that the spread remains unchanged.
Over a period $\Delta t$ this amounts to
\begin{equation}
c'\,\Delta t + (\sbar-c') \big(\Pi(T) - \Pi(T-\Delta t)\big)
\label{eq:exret_cy}
\end{equation}
and the last term is the pull-to-par. Hence the carry is not simply $\sbar\,\Delta t$. We said earlier that this calculation, despite being superficially trivial, requires a model curve, and now the reason becomes clear: we need $\Pi(T)$.

\item
Rolldown is the effect of the spread changing from rolling down the model curve.  This is the spread change multiplied by the RPV01 on the future date, assuming that all curves are unchanged:
\begin{equation}
\big(\shat(T) - \shat(T-\Delta t) \big) \Pi(T-\Delta t).
\label{eq:exret_rd}
\end{equation}

\item 
RV is the effect of moving towards the model curve at the \emph{end} of the time period, and so is
\begin{equation}
\big(\sbar - \shat(T)\big)\Pi(T-\Delta t).
\label{eq:exret_rv}
\end{equation}
The expression uses $\shat(T)$ not $\shat(T-\Delta t)$, as the latter would incorrectly double-count the rolldown effect.

\item 
Total return is the sum of all three and hence is
\begin{equation}
c'\,\Delta t + (\sbar-c')\Pi(T) - \big(\shat(T-\Delta t)-c'\big) \Pi(T-\Delta t).
\label{eq:exret_all}
\end{equation}
Notice that there is a delicate cancellation of terms en route when the components are added, giving this neat result.

\end{itemize}

In fact, the end result can be derived in a different way, at least for CDS, which makes it clear that it is correct.
Consider the PL arising from selling protection (today) when the traded spread is $\widetilde{s}_0$ and unwinding it later when it is $\widetilde{s}_1$. As usual the coupon is $c$. This is\footnote{Recall that $\widetilde{\Pi}$ is the ISDA `flat-hazard-curve' RPV01.}
\begin{equation}
c'\, \Delta t + (\widetilde{s}_0 - c) \widetilde{\Pi}(T) - (\widetilde{s}_1 - c) \widetilde{\Pi}(T-\Delta t).
\label{eq:exret2}
\end{equation}
Earlier we showed that the par spread $\sbar$ is related to the traded spread $\widetilde{s}$ by $(\sbar - c)\Pi = (\widetilde{s}-c)\widetilde{\Pi}$. 
So (\ref{eq:exret2}) corresponds exactly to (\ref{eq:exret_all}), the idea being that the par spread is $\sbar$ today and moves to the value on the model curve, which is $\shat(T-\Delta t)$, over the period $[0,\Delta t]$.

There is a different way of doing it, giving the same total return as before, but subdividing differently:
\begin{itemize}
\item
Carry differs by using the model spread $\shat=\shat(T)$:
\begin{equation}
c'\,\Delta t + (\shat-c') \big(\Pi(T) - \Pi(T-\Delta t)\big)
\tag{\ref{eq:exret_cy}a}
\label{eq:exret_cy2}
\end{equation}
\item
Rolldown is the same as above, (\ref{eq:exret_rd}).
\item
RV is the effect of moving towards the model curve at the \emph{beginning} of the time period, and so is
\begin{equation}
\big(\sbar - \shat(T)\big)\Pi(T).
\tag{\ref{eq:exret_rv}a}
\label{eq:exret_rv2}
\end{equation}
\end{itemize}
It is easily seen that the sum of (\ref{eq:exret_cy2},\ref{eq:exret_rd},\ref{eq:exret_rv2}) is the same as that of (\ref{eq:exret_cy},\ref{eq:exret_rd},\ref{eq:exret_rv}), i.e.~(\ref{eq:exret_all}).

We have talked about total return, rather than expected return. This is because the above development has assumed that the curve remains unchanged and the bond spread moves towards it over the time $[0,\delta T]$. A theory of expected return must capture more than this, as there are other considerations:
\begin{itemize}
\item
The bond may default or credit migration may occur to a different rating state. But as we have curves for each rating, this is easily captured by summing over all rating transitions, with transition probabilities taken from the appropriate row of the chosen transition matrix, for which see \cite{Martin21a}. It is a reasonable way of thinking about the probability of large and sudden price changes.
\item
Even if no transition occurs, the bond may not converge all the way to the curve off which we think it should trade. This can be captured by multiplying the RV component by a number between 0 and 1. But if this is done the two total return calculations given above, i.e.\ (\ref{eq:exret_cy2},\ref{eq:exret_rd},\ref{eq:exret_rv2}) vs (\ref{eq:exret_cy},\ref{eq:exret_rd},\ref{eq:exret_rv}), will no longer be the same.
\item
Curve dynamics. This will be pursued in further work.
\end{itemize}

\subsection{Extension to EM bonds}

An effect which has prevailed in markets for many years has been that EM issuers of the same standalone credit quality tend to trade wider than DM ones (at the same maturity point). The wider the sovereign spread, the more pronounced is the difference.

We need to grasp an important but subtle problem. When we are fitting bonds to `rating curves', exactly what do we mean by rating? The simplest treatment is the external rating, which we are at liberty to adjust if we consider it to be wrong. The problem is that that rating incoporates both the issuer's fundamentals and the sovereign. For example, suppose XYZ has credit fundamentals that would point to a A$-$ rating if it were in the US. If in Mexico (rating BBB) it is likely to be rated no higher than BBB+/BBB. Suppose it is actually rated BBB. Consider now how it is going to trade. There are several ways of thinking about it:
\begin{itemize}
\item[I] It is a standalone A$-$ but in a BBB country and we adjust the DM A$-$ curve wider to take into account the BBB sovereign;
\item[II] It is actually rated BBB (composite rating), so we just use the DM BBB curve;
\item[III] Although the external rating of BBB should take the sovereign into account, the market consistently trades such names slightly wider than their DM BBB counterparts. So the DM BBB curve needs adjusting wider, but not as much as the A$-$ curve does in (I) above.
\end{itemize}
The position is no easier for internal ratings, because we need to be clear what our internal rating means: standalone, or composite (=incorporating the sovereign to a degree that requires more precision than current practice).

An obvious advantage of (I) is that the mapping from fundamental factors (debt:EBITDA, variability of OCF, etc) to standalone rating is the same for EM as for DM, whereas producing a composite rating is less well-defined.

A recent problem in Argentina (late 2023/early 2024) reveals a problem with (II,III). All the corporates are rated CCC$-$, as that is the sovereign rating; but not all are trading at distressed levels. In this case, therefore, the combined rating results in the spread being grossly overestimated, and in fact it has no information content at all. Really, in this case the question is to work out how coupled to the sovereign, and to a collapse in the currency, each firm is, and rate them accordingly: one obvious matter is what their assets are (USD, pesos, etc., rather than just the USD equivalent, as exchange controls will forbid interconversion). On the basis of this example alone, which admittedly is a tiny fraction of the universe, (I) looks like the right idea. 

Another piece of evidence is worth bringing up. Corporates can trade tighter than their sovereigns. 
Related to this, a simple addition of standalone corporate spread to sovereign spread produces a combined spread that is far too high. In spread terms, the adjustment referred to in (I,III) above cannot be the full sovereign spread. See \cite{Martin18c}: this models the default event of a corporate as the first passage time below a barrier of either the standalone corporate or of the sovereign, where the sovereign barrier is moved downwards so that only very severe sovereign default triggers insolvency of the corporate. It is suggested in \cite{Martin18c} that the adjustment is bigger for poorer-quality corporates: hypothetically a AA is a AA regardless of jurisdiction, whereas a B is much more vulnerable.

Following on from \cite{Martin18c}, which is cumbersome to implement, we might be abe to pursue (I) by combining the survival curves of the standalone corporate and sovereign using some simple algebraic procedure. Obviously this has to generate a valid curve and have sensible properties.

Here are two ideas for forming the composite survival curve $\widetilde{Q}$:
\begin{eqnarray*}
\widetilde{Q}(T) &=& Q(T) Q_\textrm{sov}(T) \\
\widetilde{Q}(T) &=& Q(T) \big[ Q(T) + Q_\textrm{sov}(T) - Q(T)Q_\textrm{sov}(T) \big]
\end{eqnarray*}
The first is a simple first-to-default idea, and is very naive: it assumes no correlation between sovereign and standalone corporate, and assumes that default of the sovereign triggers default of the corporate. It adds the constituent IFHRs, which roughly corresponds to adding the spreads (not exactly, as they recovery rates may differ), and we know that this increases the spread too much. Nonetheless it is a useful building block.
The second is more subtle, and reduces the survival probability more for lower-grade firms, as can be seen from the fact that $Q(T)=1 \Rightarrow \widetilde{Q}(T)=1$. It is not difficult to show that it generates a valid survival curve.

By taking a convex combination of these two and the trivial model $\widetilde{Q}(T)=Q(T)$ we arrive at
\begin{equation}
\widetilde{Q}(T)= Q(T)\big[ 1-\beta-\beta' + \beta Q_\textrm{sov}(T) + \beta' (Q(T) + Q_\textrm{sov}(T) - Q(T)Q_\textrm{sov}(T))  \big]
\label{eq:combin}
\end{equation}
where we need $\beta,\beta'\ge0$, $\beta+\beta'\le1$. When $\beta=\beta'=0$ the sovereign exerts no effect. Raising either parameter causes more coupling from sovereign to corporate, with $\beta'$ having much less effect on high-grade credits than $\beta$.

\notthis{
One way of analysing it is to incorporate it into the multi-curve fitting method in a rather obvious way, altering eq.(\ref{eq:resid}) to
\begin{equation}
\Delta P_j/100 = 
\big( 1-P_j/100 + c_j-\rhat_{j,\ell(j)}(T_j)-s_{\ell(j)}(T_j) -\alpha s^\textrm{sov}_j(T_j) \big) \Pi_{\ell(j)}(T_j)
\label{eq:residem} ;
\end{equation}
where $s^\textrm{sov}(T)$ is the par CDS spread of the sovereign at the associated maturity point and the subscript $_j$ indexes the bond and $_{\ell(j)}$ its corresponding curve. The coefficient $\alpha$ must be constrained to lie between 0 and 1, and typically one finds, on fitting the curves and $\alpha$, that $\alpha$ is a little below 0.5 on average. A refinement is to notice that highly-rated EM bonds have less connection to the sovereign spread than lower-rated ones, and this effect is easily incorporated into the model.

There is a distinction between this and the work presented in \cite{Martin18c}, which attempts a more fundamental approach based on treating the EM bond as a modified first-to-default basket of a standalone issuer and its sovereign. There, the inputs to the model are the standalone rating (as determined by fundamental analysis) and the sovereign spread; here they are the combined rating (which is the external rating) and sovereign spread.

} 

\subsection{Examples and discussion}

Our original questions A,B can now be answered and we show sample results by means of figures.

\begin{figure}[!h]
\hspace{-20mm}\begin{tabular}{ll}
\scalebox{0.75}{\includegraphics{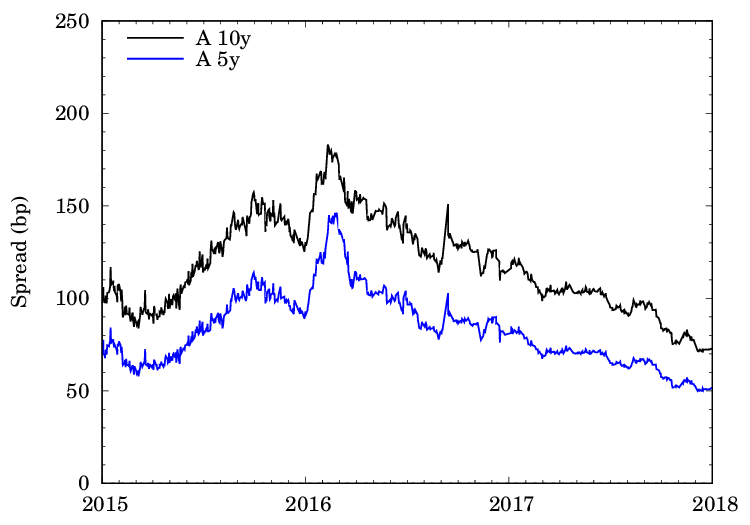}} &
\scalebox{0.75}{\includegraphics{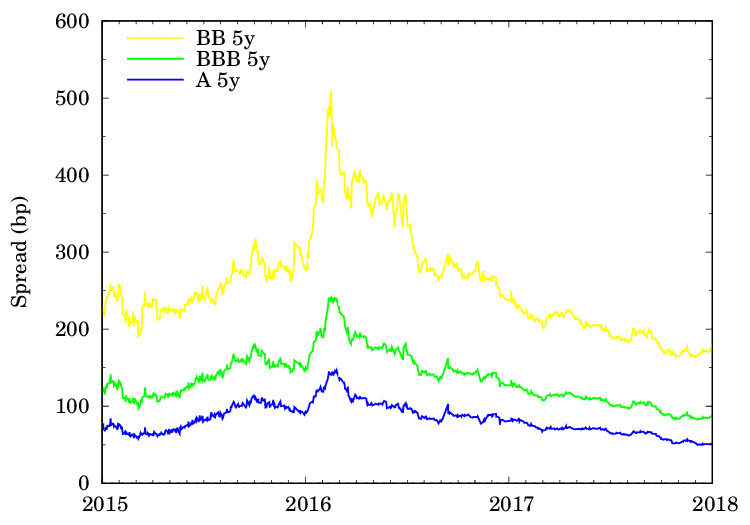}} 
\\
(i) & (ii) 
\end{tabular}
\mycaption{EM banks curve (adjusted by sovereign), 2015--2018. (i) Different points on A~curve. (ii) 5y points on different curves.}
\label{fig:embankshist}
\end{figure}

\begin{figure}[!h]
\hspace{-20mm}\begin{tabular}{ll}
\scalebox{0.75}{\includegraphics{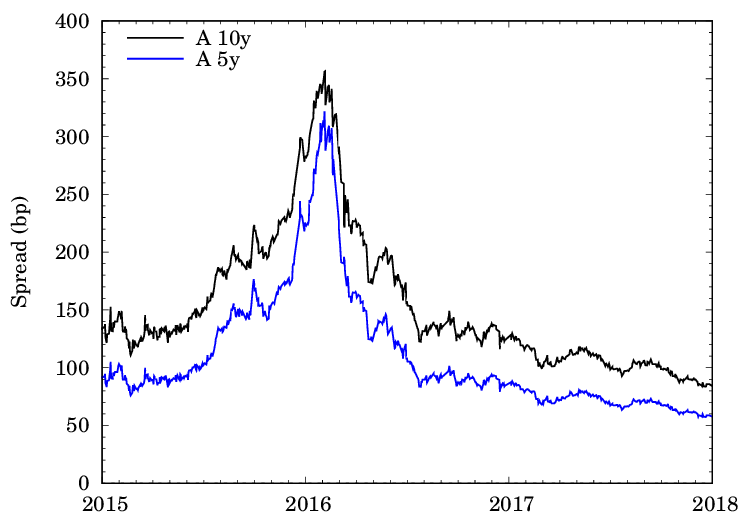}} &
\scalebox{0.75}{\includegraphics{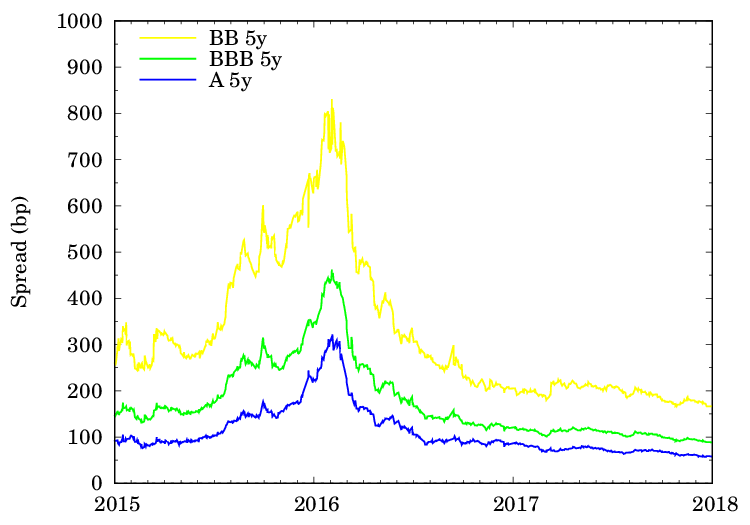}} 
\\
(i) & (ii) 
\end{tabular}
\mycaption{Mining curve, 2015--2018. (i) Different points on A~curve. (ii) 5y points on different curves.}
\label{fig:mininghist}
\end{figure}

First, Q.~A1/2/3: how has a particular curve changed over time? Figure~\ref{fig:embankshist}(i) shows, for EM banks, the 5y and 10y points on the A~rated curve over time, and hence the level and steepness, while (ii) shows the 5y point for different ratings.
As a second example, Figure~\ref{fig:mininghist} shows the mining sector. Notice how the curve flattened during the 2016 commodities crisis.

\begin{figure}[!h]
\centering
\scalebox{0.8}{\includegraphics{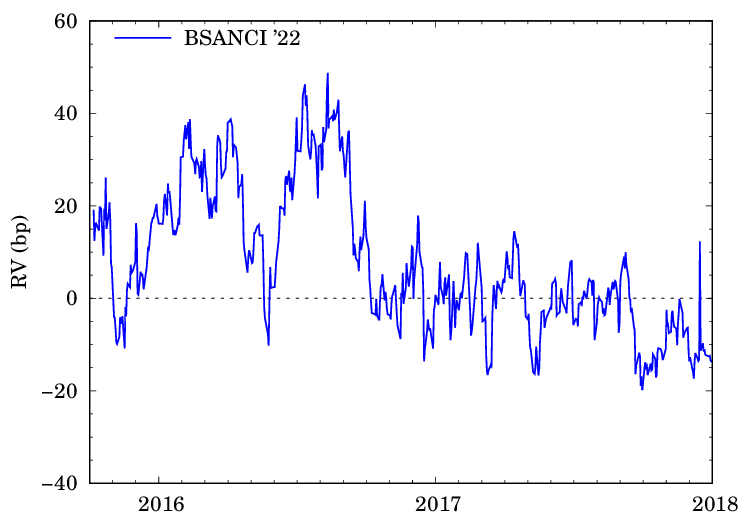}} \\
\mycaption{BSANCI~'22 relative value (+ means cheap) to the A+ banks curve, over a period of a few years.}
\label{fig:bsanci22}
\end{figure}

Next, Q.~B1: how rich or cheap is a bond to its curve? Figure~\ref{fig:bsanci22} shows the RV for BSANCI~'22 (a Chilean bank) over a period of a few years. Notice after some initial volatility, when spreads were higher, how the relative value settles down to quite a tight range about zero, as might reasonably be expected.

\clearpage

\subsection{Bond forwards}

The following is a precursor to forthcoming work on callable bonds. A general principle with anything option-like is that we need a forward and a volatility. Callable bonds are more complicated than that because they have multiple exercise dates, but we can already deal with the simplest part of the problem, i.e.\ the forward, so it is natural to do so here.

\notthis{
A callable bond is one in which the issuer has the right to buy back the bond issue at a specified price, always $\ge100$, on or after a specified date. Typically these bonds have multiple call dates, which are referred to as a schedule.
The callable bond therefore consists of a hypothetical regular (uncallable) bond with a short option position embedded in it. We said that the regular bond is purely hypothetical because it does not exist on its own: in other words, all we have is the callable bond, and maybe some other regular ones that were issued at different times.
The objective is to value the call option.
}

First we define the $T_e$-maturity \emph{conditional forward} of a bond, which is the fair value at which the bond should be agreed to be purchased at time $T_e$ from now, conditionally on no default: this last condition means that should the issuer default before $T_e$ then the contract is void. The formula is
\begin{equation}
F/100 = \frac{ P/100 - c \Pi(T_e) - \recov\, \Xi(T_e)}{ B(T_e) Q(T_e) }
\label{eq:fwd}
\end{equation}
which bears a strong resemblance to (\ref{eq:bp}) (indeed, that equation shows immediately that $F=100$ when $T_e=T$, as it must); it is understood by arbitrage arguments as follows.

Take the bond and synthetically\footnote{In other words do a total return swap on the coupon stream. In particular, if the issuer defaults then this stops paying, which is why the formula is $c\times$ the risky PV01, not the riskfree PV01.} sell off the coupon stream up to time $T_e$, thereby realising $c \Pi(T_e)$ upfront; also synthetically sell off the right to receive any recovery should default occur before $T_e$, realising $\recov\, \Xi(T_e)$ upfront.
Write $F$ for the forward price, and sell the forward contract, thereby guaranteeing to receive $F$ at time $T_e$ but only if the issuer has survived: in the event of default there will be no payment. We must therefore buy protection on a notional of $F/100$, but it needs to be of non-standard type in three respects: first, it is zero-recovery; secondly, it pays out at time $T_e$, not at the default time; thirdly, we ask to pay for it upfront. Then the cost is $(F/100) B(T_e)\big(1-Q(T_e)\big)$. It now does not matter whether the issuer defaults or not: if it does, then we deliver the bond into the CDS and receive $F/100$, whereas if it does not, we deliver it into the forward and receive the same. Now matching the time-$T_e$ payments against the upfront payments, we obtain (\ref{eq:fwd}).
The equation is of the familiar type: that is to say, spot price of instrument, less coupons/dividends, with interest added on to reflect the timing difference. The final division by the survival probability comes from the `knockout property', i.e.\ it is the conditional forward.

The \emph{unconditional forward}, i.e.\ contract that must be honoured whether default occurs or not, is the same except that $Q(T_e)$ in the denominator is removed.


\section*{Conclusions and further thoughts}

We have shown how to fit a survival curve to bond/CDS data, while avoiding the use of Z-spread, which as we have explained does not deal with non-par bonds properly. We have deliberately chosen a parsimonious model, and have found that keeping the number of parameters low gives robust fitting.

Although the following goes beyond the scope of the current paper, it is appropriate to remark on the general statistical properties of curves and relative value. It is easy to fall into the trap of making unwarranted generalisations, but the following principles should be borne in mind:
\begin{itemize}
\item
Sector spreads, and the market as a whole, typically exhibit momentum in the short to medium term---as evidenced by long rallies after crises small and large---but long-term mean reversion. The latter effect is explained by market risk premium being assumed to revert, and is evident from graphs of bond spreads over a couple of decades.
\item
The issuer RV `mean-reverts until it doesn't'. A standard pastime of bond investors is buying bonds they consider `cheap to the curve' and selling those that are `rich'. The reason that mean reversion does not occur immediately is that there is no consensus as to where the curve is at any moment; the work here shows how to do it in a way that we consider to be superior. This method of investing works well until an accident befalls a particular credit. Then it will trade wider and wider and always appear cheap, as the rating typically moves later than the price. When this happens, the RV is likely to exhibit short-term momentum, as the move away from its original curve tends not to occur in one big jump. Accordingly, the role of a fundamental analyst is to determine, when an issuer starts to trade cheap, whether this is transitory or instead because of some irreparable problem. Any competent analyst, trader or PM should be able to come up with plenty of examples of both. 
\end{itemize}

\bibliographystyle{plain}
\bibliography{}

\end{document}